\documentclass[twocolumn]{aastex631}

\usepackage[caption = false]{subfig}
\usepackage{csquotes}

\usepackage[normalem]{ulem}
\usepackage{xcolor}
\begin{document}

\title{Large Scale Linear Magnetic Holes with Magnetic Mirror Properties in Hybrid Simulations of Solar Wind Turbulence}

\author{Giuseppe Arrò}
\affiliation{Los Alamos National Laboratory, Los Alamos, NM 87545, USA}

\author{Francesco Califano}
\affiliation{Dipartimento di Fisica \enquote{E. Fermi}, Università di Pisa, Pisa, Italy}

\author{Francesco Pucci}
\affiliation{Istituto per la Scienza e Tecnologia dei Plasmi, Consiglio Nazionale delle Ricerche, Bari, Italy}

\author{Tomas Karlsson}
\affiliation{Division of Space and Plasma Physics, School of Electric Engineering and Computer Science, KTH Royal Institute of Technology, Stockholm, Sweden}

\author{Hui Li}
\affiliation{Los Alamos National Laboratory, Los Alamos, NM 87545, USA}

\begin{abstract}

Magnetic holes (MHs) are coherent magnetic field dips whose size ranges from fluid to kinetic scale, ubiquitously observed in the heliosphere and in planetary environments. Despite the longstanding effort in interpreting the abundance of observations, the origin and properties of MHs are still debated. In this letter, we investigate the interplay between plasma turbulence and MHs, using a 2D hybrid simulation initialized with solar wind parameters. We show that fully developed turbulence exhibits localized elongated magnetic depressions, whose properties are consistent with linear MHs frequently encountered in space. The observed MHs develop self-consistently from the initial magnetic field perturbations, by trapping hot ions with large pitch angles. Ion trapping produces an enhanced perpendicular temperature anysotropy that makes MHs stable for hundreds of ion gyroperiods, despite the surrounding turbulence. We introduce a new quantity, based on local magnetic field and ion temperature values, to measure the efficiency of ion trapping, with potential applications to the detection of MHs in satellite measurements. We complement this method by analyzing the ion velocity distribution functions inside MHs. Our diagnostics reveal the presence of trapped gyrotropic ion populations, whose velocity distribution is consistent with a loss cone, as expected for the motion of particles inside a magnetic mirror. Our results have potential implications for the theoretical and numerical modelling of MHs. 

\end{abstract}


\section{Introduction} 

Magnetic holes (MHs) are coherent magnetic field depressions observed ubiquitously in space. Satellite measurements have revealed the presence of MHs in the solar wind (SW) \citep{turner1977magnetic,winterhalter2000latitudinal,perrone2016compressive,yu2021characteristics}, planetary magnetosheaths \citep{volwerk2008mirror,genot2009mirror,madanian2020magnetic,karlsson2021magnetic,huang2021situ}, in the Earth's magnetotail \citep{huang2019mms}, and in cometary environments \citep{russell1987mirror}. The size of MHs is variable, ranging from hundreds of ion gyroradii $\rho_i$, to a few electron gyroradii $\rho_e$ \citep{stevens2007scale}. Henceforth, we will call \enquote{large scale MHs}, those MHs with size of about $10\,\rho_i$ or more. Magnetic holes are typically associated with local density and temperature enhancements, roughly balancing the magnetic pressure drop. The temperature increase is not isotropic, with higher temperatures perpendicular to the local magnetic field. The anticorrelation between the magnetic field and the density, together with the enhanced perpendicular temperature anisotropy, typically exceeding the mirror instability threshold, are all features hinting at a connection between MHs and mirror modes \citep{pantellini1998model}. However, it is still unclear whether the mirror instability is actually capable of generating MHs. This problem has been investigated numerically by \citet{califano2008nonlinear} and \citet{shoji2012multidimensional}, using hybrid simulations. Their studies have shown that the nonlinear stage of the mirror instability consists of a sequence of magnetic peaks and dips that merge over time, producing mainly magnetic peaks, while holes develop only under very specific circumstances, rarely consistent with observations. Therefore, \citet{califano2008nonlinear} proposed that MHs may actually be a stable solution of the Vlasov-Maxwell system rather than a byproduct of the mirror instability. According to satellite measurements, mirror structures consisting of sequential peaks and dips, and isolated magnetic peaks, are generally observed in mirror-unstable environments \citep{soucek2008properties}. Conversely, MHs are most often observed as isolated structures in mirror-stable regions \citep{balikhin2009themis}. Hence, the occurrence of isolated MHs is hardly explained by models based on the mirror instability \citep{sundberg2015properties}.

Turbulence has also been suggested as a possible driver for the formation of MHs, as the latter are often found in turbulent environments \citep{huang2017magnetospheric}. This idea is corroborated by fully kinetic numerical simulations, showing that turbulent plasmas produce sub-ion scale MHs \citep{haynes2015electron,roytershteyn2015generation,arro2023generation}. The interplay between plasma turbulence and the mirror instability has been investigated numerically by \citet{hellinger2017mirror}, where a mirror-unstable plasma subject to turbulent perturbations has been considered. This work has shown that even in the presence of turbulence, a mirror-unstable plasma ultimately produces isolated magnetic peaks, rather than MHs.

In this letter, we investigate the formation and properties of large scale MHs in a 2D hybrid simulation of plasma turbulence, starting from an initially mirror-stable plasma. We show an alternative mechanism for the generation of MHs, where strong magnetic fluctuations stabilize by trapping ions, eventually relaxing into long lived MHs that coexist with fully developed turbulence. Despite the similarities with mirror modes, these stable MHs are not produced by the mirror instability, and their temperature anisotropy develops as a stabilizing effect, rather than being the source of free energy for their formation. 

\section{Simulation setup}

\begin{figure*}[t]
\centering
\subfloat{
\includegraphics[width=0.61\linewidth]{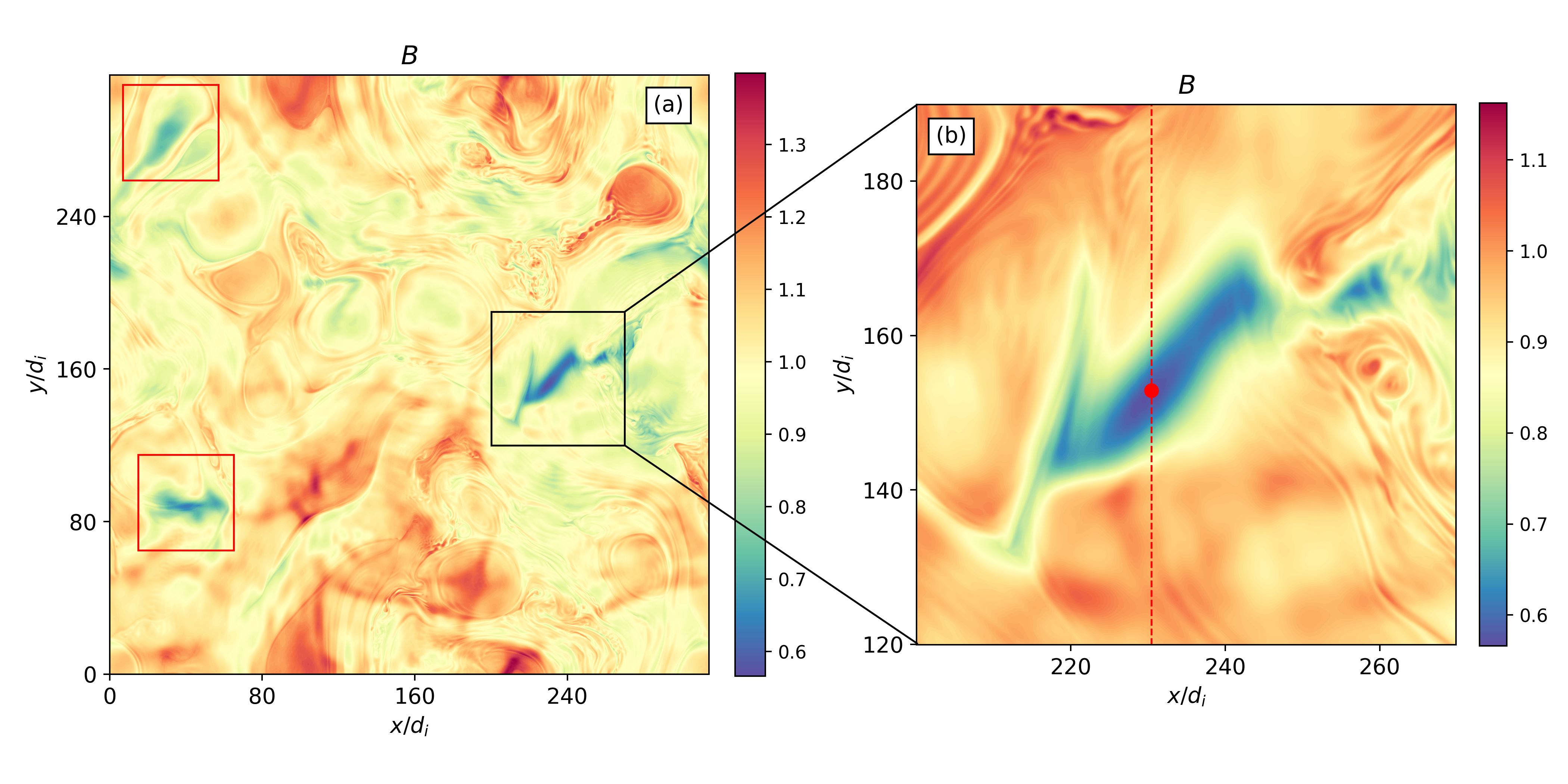}
}
\subfloat{
\includegraphics[width=0.35\linewidth]{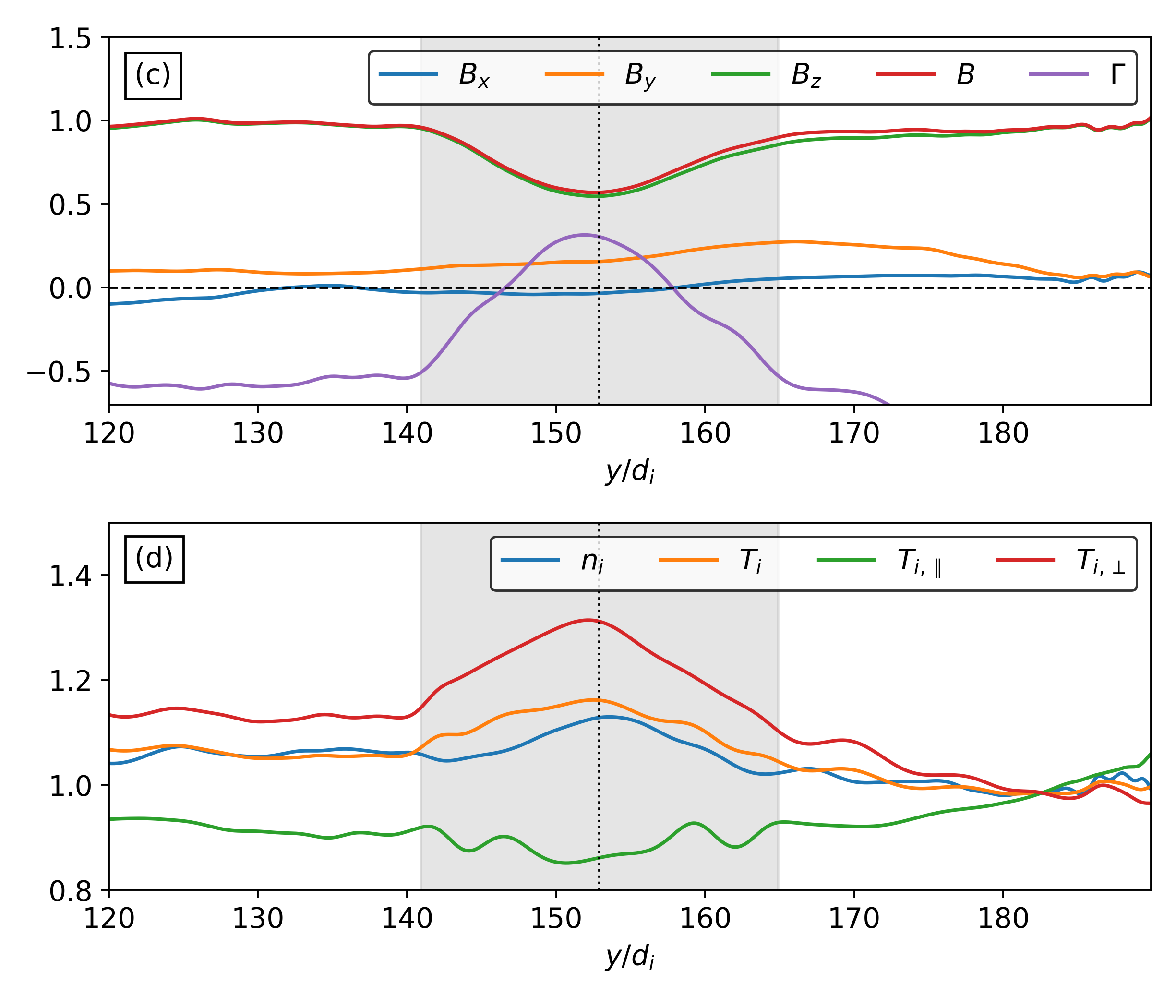}
}
\vspace{-0.5cm}
\subfloat{
\includegraphics[width=0.31\linewidth]{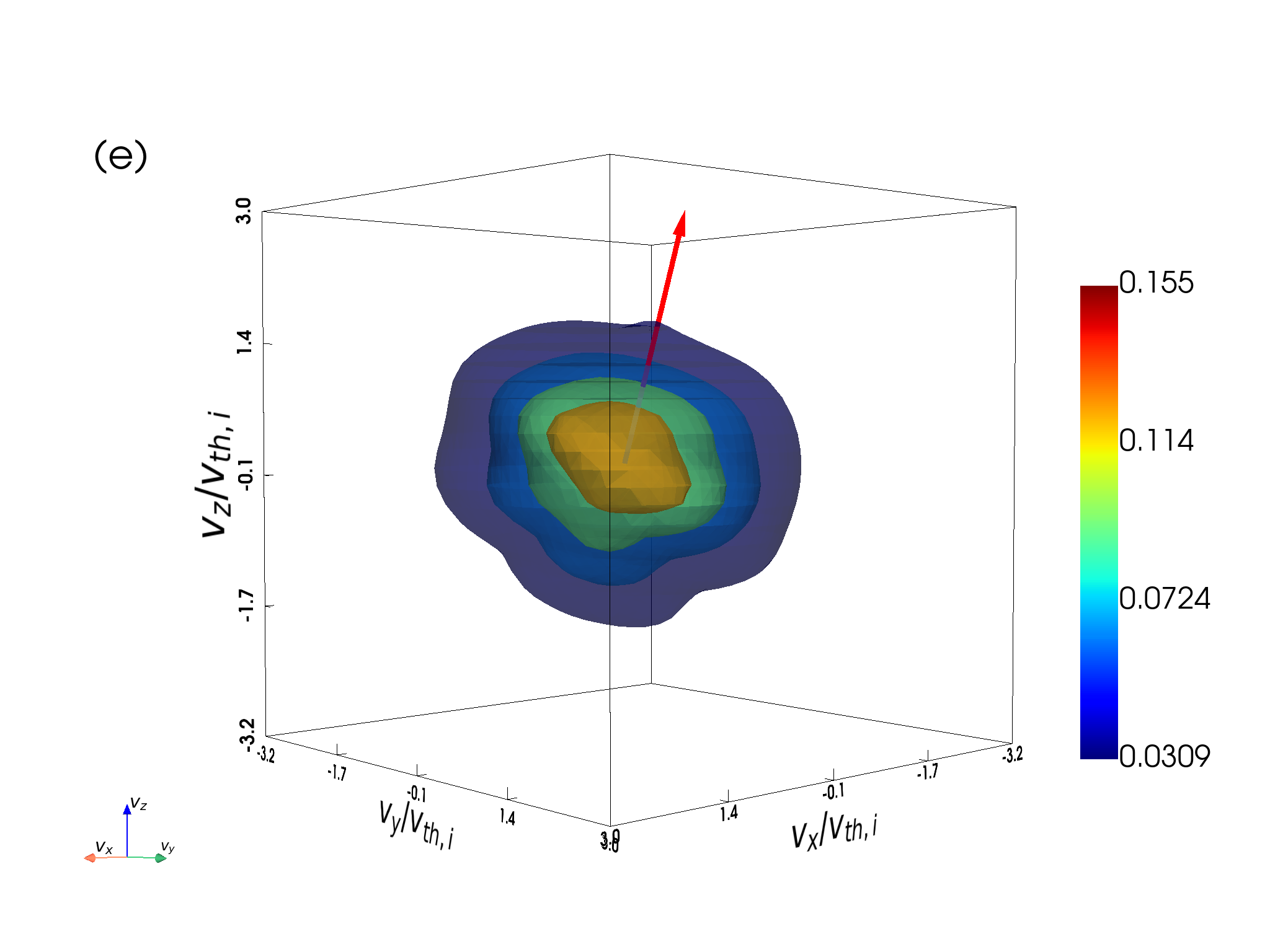}
}
\subfloat{
\includegraphics[width=0.31\linewidth]{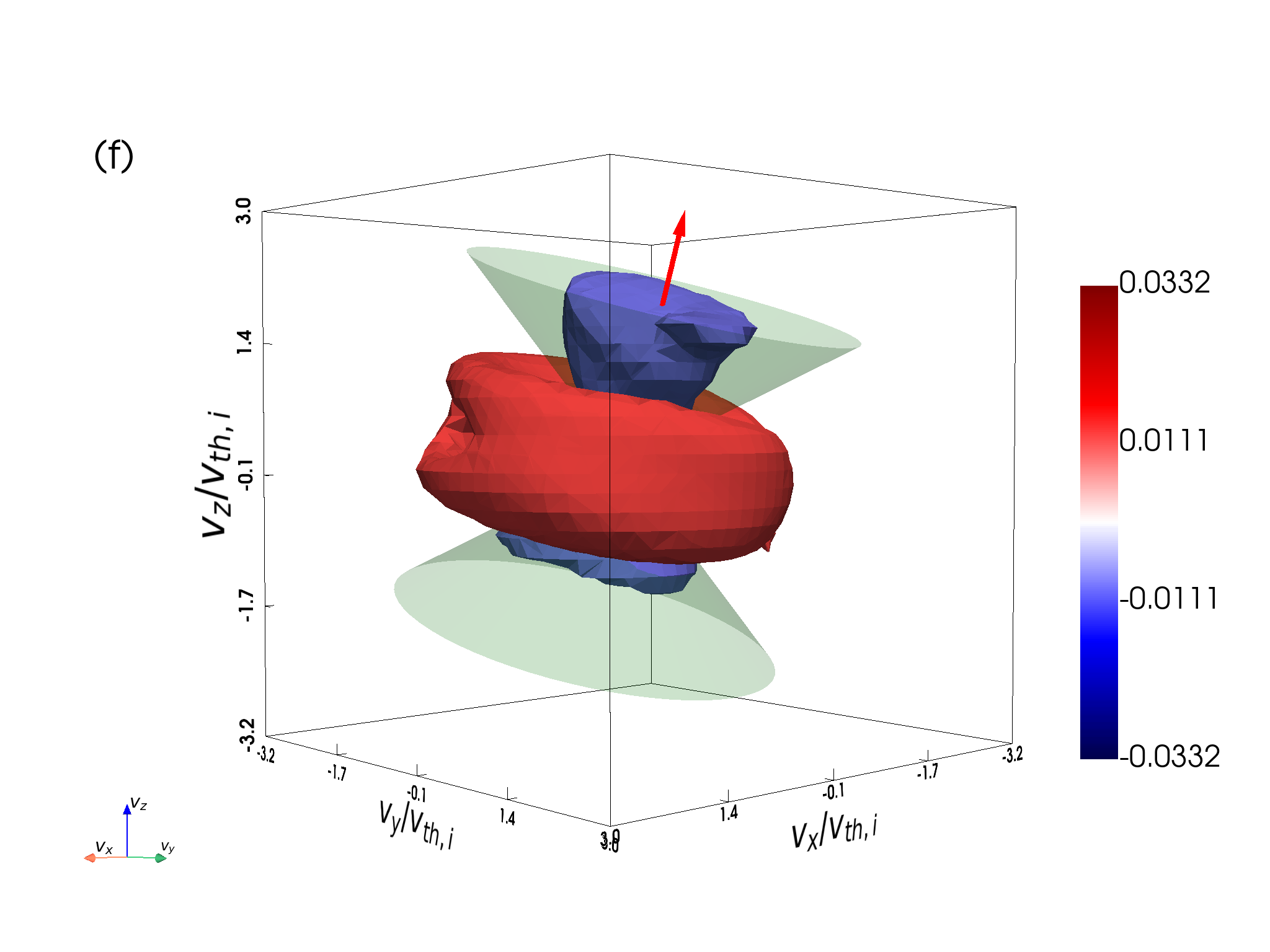}
}
\subfloat{
\includegraphics[width=0.34\linewidth]{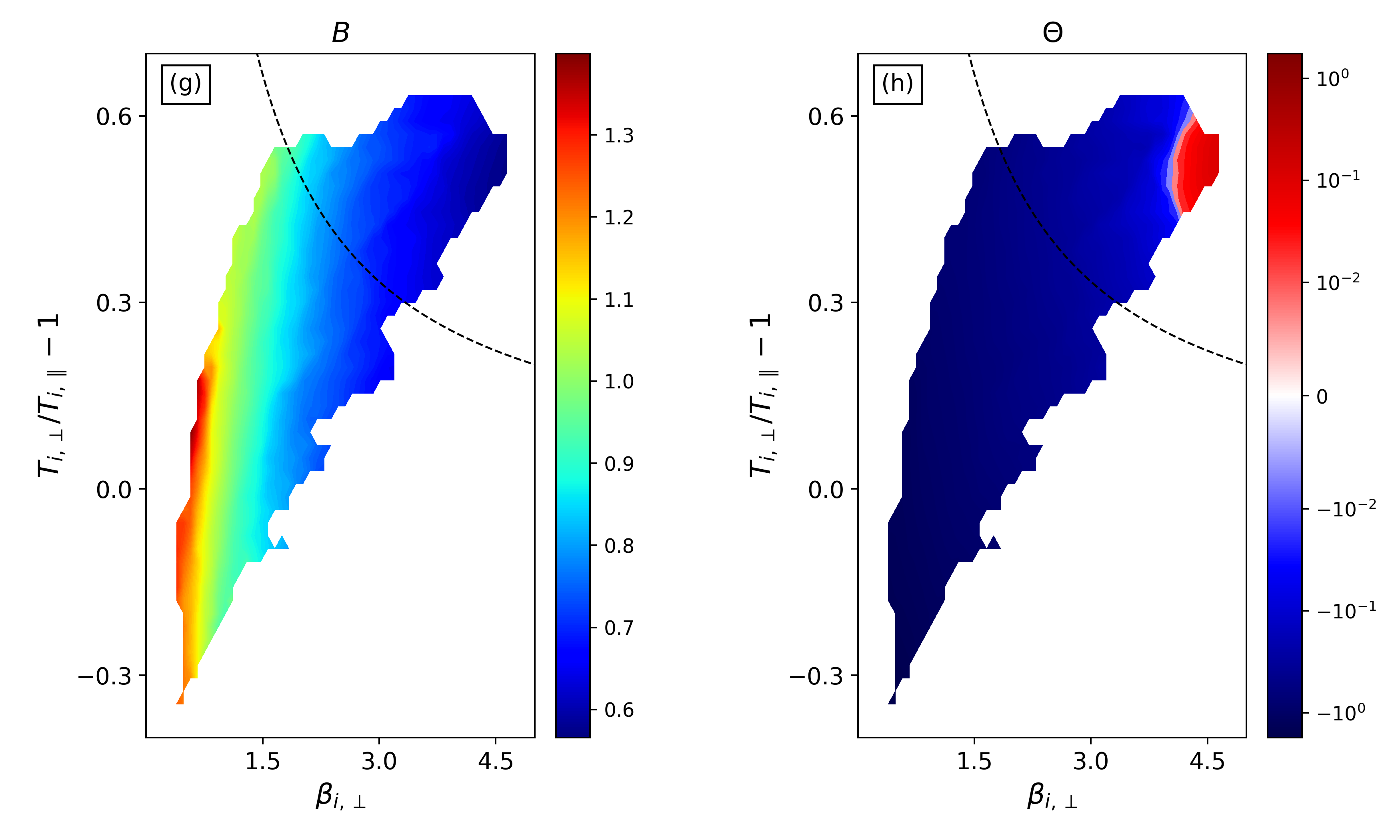}
}
\caption{Magnetic field magnitude at $t\!=\!543.5\,\Omega_i^{-1}$ (a), with a zoom into a MH (b); Magnetic field components and mirror threshold $\Gamma$ (c), ion density and temperature components (d), over the red dashed line in panel (b); 3D isosurfaces of $f_i$ (e) and $\delta f_i$ (f), at the red dot in panel (b), with local magnetic field (red arrow) and loss cone (green surface); Magnetic field magnitude (g) and $\Theta$ (h) distributions in the $(T_{i,\perp}/T_{i,\parallel}-1,\,\beta_{i,\perp})$ plane, with mirror threshold (black dashed line). The magnetic field, density and temperature are in units of $B_0$, $n_0$ and $T_0$, respectively.}
\label{MH}
\end{figure*}

Our simulation has been realized using the Eulerian Hybrid Vlasov-Maxwell (HVM) code \citep{valentini2007hybrid}, with kinetic ions and fluid isothermal electrons with finite mass. The spatial domain is a 2D square periodic uniform grid with $3072^2$ points, and size $L\!=\!100\pi\,d_i$, where $d_i$ is the ion inertial length. The velocity domain is a 3D uniform grid with $51^3$ points, with ion velocities $v_x,\,v_y,\,v_z$ in the range $-5\,v_{th,i}\!\leqslant\! v_x,\,v_y,\,v_z \!\leqslant\!+5\,v_{th,i}$, where $v_{th,i}$ is the initial ion thermal speed. Ions are initially Maxwellian, with uniform density $n_0$, zero mean velocity, and isotropic temperature $T_0$ with unitary ion beta $\beta_i\!=\!1$, implying $\rho_i\!\simeq\!d_i$. The ion-to-electrons mass and temperature ratios are $m_i/m_e\!=\!100$ and $T_i/T_e\!=\!1$, respectively. The initial magnetic field includes a guide field $\textbf{B}_0\!=\!B_0\hat{\textbf{z}}$, perturbed by random-phase, isotropic magnetic field fluctuations $\delta\textbf{B}$, with wavenumber $k$ in the range $1\!\leqslant\!k/k_0\!\leqslant\!6$ (with $k_0\!=\!2\pi/L$), and root mean square (rms) amplitude $\delta B_{rms}/B_0\!\simeq\!0.28$ \citep{finelli2021study}. These parameters aim at reproducing SW conditions \citep{bandyopadhyay2020situ}. No external driving is employed, making the turbulence freely decaying. The simulation time step is $dt\!=\!0.01\,\Omega^{-1}_i$ (where $\Omega_i$ is the ion cyclotron frequency), and we stop the run at $t\!=\!543.5\,\Omega_i^{-1}$, when turbulence is fully developed. We use a compact finite difference filter \citep{lele1992compact} to smooth electromagnetic fluctuations at small scales, mimicking resistivity. The smoothing does not affect ion dissipation, the latter taking place at scales of a few $d_i$, because of kinetic effects \citep{arro2020statistical,arro2022spectral}.

\section{Results}

\begin{figure*}[t]
\centering
\includegraphics[width=0.96\linewidth]{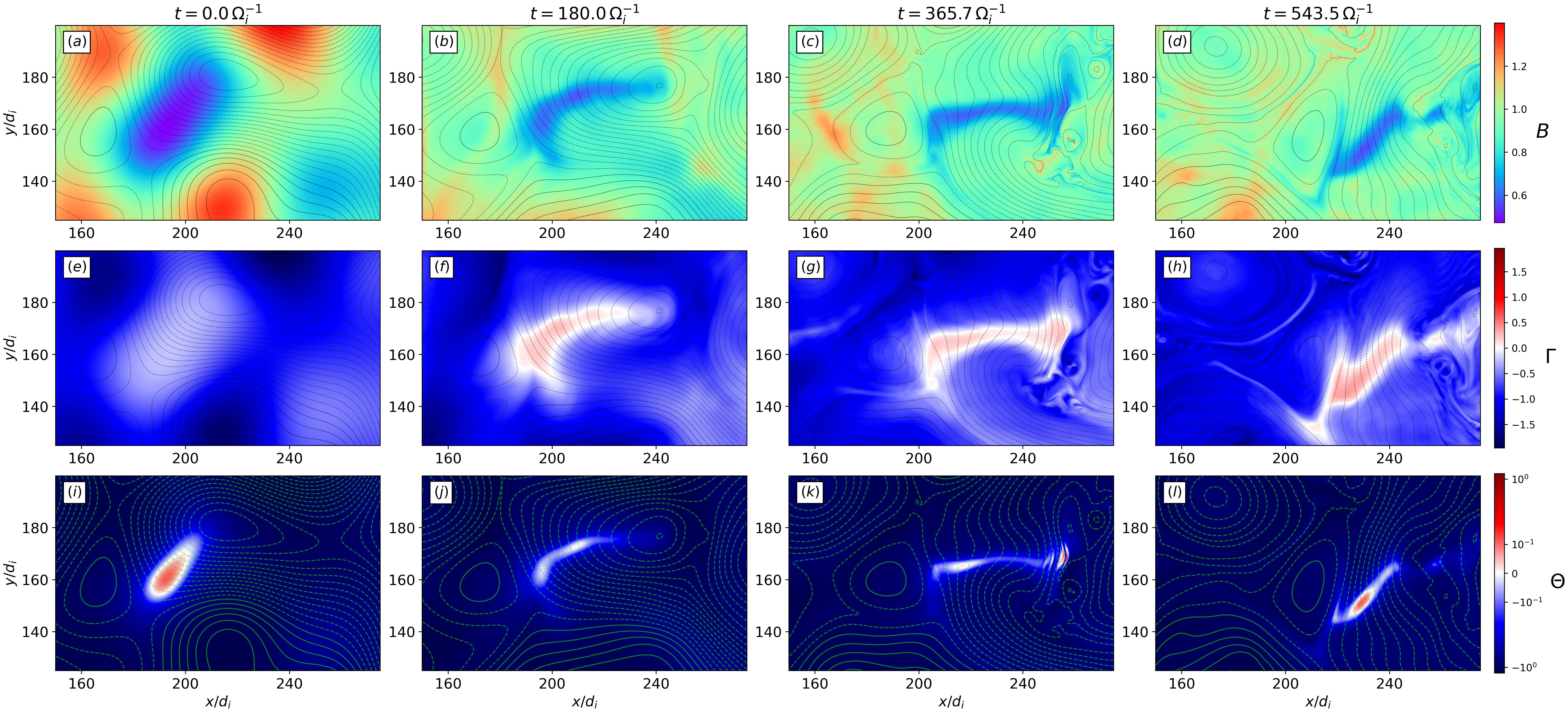}
\caption{Magnetic field magnitude (in units of $B_0$) (a-d), mirror threshold $\Gamma$ (e-h), and $\Theta$ (i-l), at different times, showing the MH formation (with isolines indicating the in-plane magnetic field). Each column corresponds to a different time, given in the column title.}
\label{time}
\end{figure*}

In the first part of this section, we will examine the properties of MHs observed in our simulation at fully developed turbulence. We will then discuss the formation and evolution of MHs, as turbulence develops. 

Panel (a) of Figure~\ref{MH} shows the magnetic field magnitude $B$ over the whole simulation domain, at fully developed turbulence ($t\!=\!543.5\,\Omega_i^{-1}$). A wide variety of structures at different scales are observed, including elongated regions where $B$ suddenly drops. Panel (b) shows a zoom into one of these magnetic dips, whose length and width are about $30\,d_i$ and $10\,d_i$, respectively. This structure has properties consistent with large scale MHs typically observed in space. Panels (c) and (d) show different quantities over a 1D cut crossing the MH, indicated by the red dashed line in panel (b). In panel (c), we see that the magnetic dip is mainly induced by the out-of-plane magnetic field $B_z$, which is dominant with respect to the in-plane components $B_x$ and $B_y$. Additionally, neither $B_x$ nor $B_y$ show strong variations correlated with the hole, implying a weak magnetic field rotation across the MH. This property is typical of the so called linear MHs (LMHs). The latter are usually observed in mirror-stable environments \citep{angeo-39-239-2021}, which is also the case in our simulation. The mirror instability threshold is \citep{southwood1993mirror}\\ 
\begin{equation}
\Gamma=\frac{T_{i,\perp}}{T_{i,\parallel}}-1-\frac{1}{\beta_{i,\perp}},    
\end{equation}
\\where $T_{i,\perp}$ and $T_{i,\parallel}$ are the perpendicular and parallel ion temperatures, with respect to the local magnetic field, and $\beta_{i,\perp}$ is the perpendicular ion beta. The threshold $\Gamma$ is rigorously valid for bi-Maxwellian ions, but it is nonetheless a useful quantity to investigate the level of temperature anisotropy, largely used in observations. Panel (c) shows that the plasma exceeds the mirror threshold ($\Gamma\!>\!0$) only inside the MH, while being mirror-stable elsewhere. The ion density $n_i$ and temperature $T_i$ have a modest increase inside the MH, as observed in panel (d). The temperature variation is mainly induced by $T_{i,\perp}$, exhibiting a significant increase, while $T_{i,\parallel}$ slightly decreases. To understand the kinetic origin of density and temperature variations, we analyze the ion velocity distribution function (IVDF) $f_i$ inside the MH. Panel (e) shows the 3D isosurfaces of $f_i$, at a position indicated by the red dot in panel (b), and by the vertical dotted line in panels (c) and (d). The IVDF is roughly gyrotropic and ions have smaller velocities along the local magnetic field (red arrow), with a wider velocity spread in the perpendicular plane. The shape of $f_i$ and the magnetic field configuration of the MH suggest that the structure may actually be able to selectively trap ions, according to their velocity, similarly to magnetic mirrors. We thus investigate the IVDF evolution by comparing $f_i$ with the initial Maxwellian distribution $g_i$. To this end, we calculate $\delta f_i\!=\!f_i-g_i$, with both $f_i$ and $g_i$ centered at the same mean velocity. A 3D representation of $\delta f_i$ is shown in panel (f), together with the local loss cone (green surface), corresponding to\\
\begin{equation}
\left| \frac{v_{\parallel}}{v_{\perp}} \right| = \sqrt{\frac{B_{out}}{B_{in}}-1} = \sqrt{\frac{B_0}{B}-1} = \frac{1}{tan(\theta_p)},  
\label{cone}
\end{equation}
\\where $\left|v_{\parallel}/v_{\perp}\right|\!$ is the parallel-to-perpendicular particle velocity ratio. $B_{in}\!=\!B$ is the magnetic field inside the MH, while $B_{out}$ is the magnetic field outside the structure, roughly equal to the guide field $B_0$, as seen in panel (c). Ions with pitch angle larger than $\theta_p$ are confined by the magnetic field configuration, while other ions escape. We see that $\delta f_i$ is positive outside the loss cone, meaning that the ion density in that portion of the velocity space has increased, because of trapped ions. On the other hand, $\delta f_i$ is negative inside the loss cone, implying that ions with small pitch angles have escaped the MH, reducing the density in the corresponding region of the velocity space. This analysis shows that the IVDF evolution is consistent with the dynamics of trapped and escaping ions, which also explains the observed temperature anisotropy inside the MH. In other words, the increase in $T_{i,\perp}$ and the slight decrease in $T_{i,\parallel}$ observed inside the MH, are caused by the fact that ions with large pitch angle are trapped by the structure, while other ions escape. The net effect of this dynamics is the development of a perpendicular temperature anisotropy above the mirror instability threshold.

Beside the MH we have analyzed so far, two more MHs with analogous properties are present in our simulation, within red boxes in panel (a) of Figure~\ref{MH}. To conveniently visualize the properties of all the MHs in the simulation, we analyze the distribution of $B$ in the $(T_{i,\perp}/T_{i,\parallel}-1$ vs $\beta_{i,\perp})$ plane. We also introduce the trapping threshold $\Theta$, to determine if magnetic structures can trap ions. $\Theta$ is derived by taking the square of Equation~\ref{cone}, which gives:\\
\begin{equation}
\Theta = \frac{B_0}{B}-1-\frac{T_{i,\parallel}}{T_{i,\perp}},   
\end{equation}
\\where we have used $\left(v_{\parallel}/v_{\perp}\right)^2\!\sim\!T_{i,\parallel}/T_{i,\perp}$, giving the average parallel-to-perpendicular ion thermal velocity ratio. $\Theta\!>\!0$ implies that ions are trapped by the local magnetic field on average. Panels (g) and (h) of Figure~\ref{MH} show the distributions of $B$ and $\Theta$, respectively. The black dashed line represents the mirror instability threshold $T_{i,\perp}/T_{i,\parallel}-1\!=\!1/\beta_{i,\perp}$. We see a monotonic tendency for regions with low $B$ to lie above the mirror threshold. Furthermore, the deepest magnetic depressions efficiently trap ions, as $\Theta\!>\!0$. Hence, MHs in our simulation behave as magnetic mirrors, trapping ions and developing a perpendicular temperature anisotropy that exceeds the mirror threshold. 

Our analysis at fully developed turbulence suggests that MHs are not produced by the mirror instability, since they are surrounded by a mirror-stable plasma, and their enhanced perpendicular temperature anisotropy develops as a consequence of ion trapping. We thus choose the MH of Figure~\ref{MH} and follow its evolution over time, to understand its generation process. Figure~\ref{time} shows $B$ (first row), $\Gamma$ (second row), and $\Theta$ (third row), around the developing MH, at different time steps (corresponding to different columns). The first column shows that a mirror-stable ($\Gamma\!<\!0$) magnetic depression is present since $t\!=\!0\,\Omega_i^{-1}$, induced by initial fluctuations. $\Theta$ is positive inside the magnetic dip, meaning it can trap ions. The second column shows that this initial magnetic dip has evolved into a MH, after $180\,\Omega_i^{-1}$. The magnetic structure has become thinner, with sharper boundaries, and the temperature anisotropy has grown above the mirror threshold ($\Gamma\!>\!0$), as a consequence of ion trapping. $\Theta$ is also positive inside the MH, but it has become more localized and weaker in amplitude. After its development, the MH remains stable and its properties are not significantly altered by the surrounding turbulence, whose main effect is to perturb the shape of the structure. By comparing the third and fourth columns of Figure~\ref{time}, we see a tendency for the MH to stay tied to a bundle of magnetic field lines, indicated by the in-plane isolines. The turbulence shakes the magnetic bundle over time, and the MH adjusts its position and shape accordingly, without noticeable variations in the values of $B$ and $\Gamma$. We note that the MH is elongated across the magnetic bundle, while being thinner along magnetic field lines. We argue that the width of the MH along magnetic field lines could be determined by the distance travelled by trapped ions, while the MH length across the magnetic bundle is roughly constant over time, likely determined by the size of the initial fluctuation. The same kind of temporal evolution and properties are observed also for the other MHs developing in the simulation.

As a final remark, we note that the MH remains stable despite its temperature anisotropy exceeding the mirror instability threshold. This is because the mirror threshold refers to a situation where linear perturbations develop over an homogeneous background, while the MH is a nonlinear structure, far from being homogeneous. Furthermore, the MH size is comparable to the typical wavelength of mirror modes, so local homogeneity is not a suitable assumption either, and linear mirror theory does not apply. From a dynamical standpoint, temperature variations inside the MH produce a pressure force that helps sustaining the magnetic depression. Thus, the enhanced temperature anisotropy plays a stabilizing effect, rather than being a source of free energy for instabilities, as confirmed by Figure~\ref{time}, where the MH appears stable for hundreds of ion gyroperiods, with no unstable wave activity observed. As can be inferred from panels (c) and (d) of Figure~\ref{MH}, the MH is not exactly in pressure balance, likely because of turbulence, which explains why the structure is stable but not stationary.

\section{Discussion and conclusions}

In this letter, we have investigated the formation and properties of large scale MHs using a 2D hybrid simulation of SW turbulence. We have shown that large amplitude magnetic fluctuations can trap ions, similarly to magnetic mirrors, evolving into MHs. Ions trapped inside MHs exhibit a velocity distribution consistent with a loss cone, and induce a strong perpendicular temperature anisotropy, above the mirror threshold, stabilizing the structure. The resulting MHs persist for hundreds of ion gyroperiods, being mildly affected by the surrounding mirror-stable turbulence. Our work shows that MHs in fully developed turbulence effectively represent a potential well that traps hot ions with large pitch angles. Thus, MHs potentially play a significant role in regulating the temperature anisotropy of turbulent plasmas such as the SW, given their abundance in these environments \citep{yu2021characteristics}. 

In this work, MHs develop from the initial magnetic field fluctuations used to drive turbulence. Initial fluctuations have an rms amplitude consistent with SW observation \citep{bandyopadhyay2020situ}, but can be quite strong locally. It is reasonable to think that strong local fluctuations could be spontaneously produced by SW turbulence, where energy cascades from scales much larger than the MHs size. Additionally, local magnetic reconnection events may also produce strong magnetic depressions \citep{pritchett2009asymmetric}. However, simulating the SW cascade from hundreds of $d_i$, retaining ion kinetic effects, is an extremely challenging problem. Hybrid particle-in-cell (PIC) codes can handle very large systems, but ion heating is sensitive to numerical noise \citep{franci2015high}. Since anisotropic ion heating is necessary to produce MHs, we employed an Eulerian Vlasov code, which is noise-free but computationally much more expensive than PIC codes. Therefore, our approach is a trade-off between numerical accuracy and a sufficiently realistic representation of SW turbulence. Nonetheless, we have shown that strong local magnetic depressions relax to stable MHs, rather than being assimilated and destroyed by turbulence. 

Understanding whether MHs in the heliosphere and planetary magnetosheaths originate in situ or come from neighbouring environments, is another open problem regarding these structures. One hypothesis is that MHs develop close to the Sun and are then advected by the turbulent SW. \citet{angeo-40-687-2022} have shown that MHs of SW origin can cross planetary bow shocks and enter the magnetosheath, where conditions for local generation are hardly met. In this context, our results show that MHs are a stable element of turbulence, supporting the idea that they can be locally generated in the SW and then transported to planets and other regions in the heliosphere. Additional factors that have not been considered in this work, such as the SW expansion and 3D effects, may also influence the stability of MHs in SW turbulence, and will be addressed in future studies.

\begin{acknowledgments}

This research was supported by the International Space Science Institute (ISSI) in Bern, through the ISSI International Team project 517: Toward a Unifying Model for Magnetic Depressions in Space Plasmas. The authors gratefully acknowledge useful discussions with all the members of the ISSI team.

This research is supported by the NASA grant No. 80NSSC23K0101. Any opinions, findings, and conclusions expressed in this work are those of the author(s) and do not necessarily reflect the views of NASA.

F. P. acknowledges support from the Research Foundation – Flanders (FWO), Junior research project on fundamental research G020224N.

Numerical simulations and data analysis have been performed on Marconi at CINECA (Italy), under the ISCRA initiative.

\end{acknowledgments}

\bibliography{MH}{}

\begin{thebibliography}{}
\expandafter\ifx\csname natexlab\endcsname\relax\def\natexlab#1{#1}\fi
\providecommand{\url}[1]{\href{#1}{#1}}
\providecommand{\dodoi}[1]{doi:~\href{http://doi.org/#1}{\nolinkurl{#1}}}
\providecommand{\doeprint}[1]{\href{http://ascl.net/#1}{\nolinkurl{http://ascl.net/#1}}}
\providecommand{\doarXiv}[1]{\href{https://arxiv.org/abs/#1}{\nolinkurl{https://arxiv.org/abs/#1}}}

\bibitem[{Arr{\`o} {et~al.}(2020)Arr{\`o}, Califano, \&
  Lapenta}]{arro2020statistical}
Arr{\`o}, G., Califano, F., \& Lapenta, G. 2020, Astronomy \& Astrophysics,
  642, A45

\bibitem[{Arr{\`o} {et~al.}(2023)Arr{\`o}, Pucci, Califano, Innocenti, \&
  Lapenta}]{arro2023generation}
Arr{\`o}, G., Pucci, F., Califano, F., Innocenti, M.~E., \& Lapenta, G. 2023,
  The Astrophysical Journal, 958, 11

\bibitem[{{Arró, G.} {et~al.}(2022){Arró, G.}, {Califano, F.}, \& {Lapenta,
  G.}}]{arro2022spectral}
{Arró, G.}, {Califano, F.}, \& {Lapenta, G.} 2022, A\&A, 668, A33,
  \dodoi{10.1051/0004-6361/202243352}

\bibitem[{Balikhin {et~al.}(2009)Balikhin, Sagdeev, Walker, Pokhotelov, Sibeck,
  Beloff, \& Dudnikova}]{balikhin2009themis}
Balikhin, M., Sagdeev, R., Walker, S., {et~al.} 2009, Geophysical Research
  Letters, 36

\bibitem[{Bandyopadhyay {et~al.}(2020)Bandyopadhyay, Sorriso-Valvo, Chasapis,
  Hellinger, Matthaeus, Verdini, Landi, Franci, Matteini, Giles,
  {et~al.}}]{bandyopadhyay2020situ}
Bandyopadhyay, R., Sorriso-Valvo, L., Chasapis, A., {et~al.} 2020, Physical
  review letters, 124, 225101

\bibitem[{Califano {et~al.}(2008)Califano, Hellinger, Kuznetsov, Passot, Sulem,
  \& Tr{\'a}vn{\'\i}{\v{c}}ek}]{califano2008nonlinear}
Califano, F., Hellinger, P., Kuznetsov, E., {et~al.} 2008, Journal of
  Geophysical Research: Space Physics, 113

\bibitem[{Finelli {et~al.}(2021)Finelli, Perri, Sisti, \&
  Califano}]{finelli2021study}
Finelli, F., Perri, S., Sisti, M., \& Califano, F. 2021, Astronomy \&
  Astrophysics, 656, A43

\bibitem[{Franci {et~al.}(2015)Franci, Landi, Matteini, Verdini, \&
  Hellinger}]{franci2015high}
Franci, L., Landi, S., Matteini, L., Verdini, A., \& Hellinger, P. 2015, The
  Astrophysical Journal, 812, 21

\bibitem[{G{\'e}not {et~al.}(2009)G{\'e}not, Budnik, Hellinger, Passot,
  Belmont, Tr{\'a}vn{\'\i}{\v{c}}ek, Sulem, Lucek, \&
  Dandouras}]{genot2009mirror}
G{\'e}not, V., Budnik, E., Hellinger, P., {et~al.} 2009, Annales Geophysicae,
  27, 601

\bibitem[{Haynes {et~al.}(2015)Haynes, Burgess, Camporeale, \&
  Sundberg}]{haynes2015electron}
Haynes, C.~T., Burgess, D., Camporeale, E., \& Sundberg, T. 2015, Physics of
  Plasmas, 22

\bibitem[{Hellinger {et~al.}(2017)Hellinger, Landi, Matteini, Verdini, \&
  Franci}]{hellinger2017mirror}
Hellinger, P., Landi, S., Matteini, L., Verdini, A., \& Franci, L. 2017, The
  Astrophysical Journal, 838, 158

\bibitem[{Huang {et~al.}(2017)Huang, Sahraoui, Yuan, He, Zhao, Le~Contel, Deng,
  Zhou, Fu, Shi, {et~al.}}]{huang2017magnetospheric}
Huang, S., Sahraoui, F., Yuan, Z., {et~al.} 2017, The Astrophysical Journal
  Letters, 836, L27

\bibitem[{Huang {et~al.}(2019)Huang, He, Yuan, Sahraoui, Le~Contel, Deng, Zhou,
  Fu, Jiang, Yu, {et~al.}}]{huang2019mms}
Huang, S., He, L., Yuan, Z., {et~al.} 2019, The Astrophysical Journal, 875, 113

\bibitem[{Huang {et~al.}(2021)Huang, Lin, Yuan, Jiang, Wei, Xu, Zhang, Zhang,
  Xiong, \& Yu}]{huang2021situ}
Huang, S., Lin, R., Yuan, Z., {et~al.} 2021, The Astrophysical Journal, 922,
  107

\bibitem[{Karlsson {et~al.}(2021)Karlsson, Heyner, Volwerk, Morooka, Plaschke,
  Goetz, \& Hadid}]{karlsson2021magnetic}
Karlsson, T., Heyner, D., Volwerk, M., {et~al.} 2021, Journal of Geophysical
  Research: Space Physics, 126, e2020JA028961

\bibitem[{Karlsson {et~al.}(2022)Karlsson, Trollvik, Raptis, Nilsson, \&
  Madanian}]{angeo-40-687-2022}
Karlsson, T., Trollvik, H., Raptis, S., Nilsson, H., \& Madanian, H. 2022,
  Annales Geophysicae, 40, 687, \dodoi{10.5194/angeo-40-687-2022}

\bibitem[{Lele(1992)}]{lele1992compact}
Lele, S.~K. 1992, Journal of computational physics, 103, 16

\bibitem[{Madanian {et~al.}(2020)Madanian, Halekas, Mazelle, Omidi, Espley,
  Mitchell, \& McFadden}]{madanian2020magnetic}
Madanian, H., Halekas, J., Mazelle, C., {et~al.} 2020, Journal of Geophysical
  Research: Space Physics, 125, e2019JA027198

\bibitem[{Pantellini(1998)}]{pantellini1998model}
Pantellini, F.~G. 1998, Journal of Geophysical Research: Space Physics, 103,
  4789

\bibitem[{Perrone {et~al.}(2016)Perrone, Alexandrova, Mangeney, Maksimovic,
  Lacombe, Rakoto, Kasper, \& Jovanovic}]{perrone2016compressive}
Perrone, D., Alexandrova, O., Mangeney, A., {et~al.} 2016, The Astrophysical
  Journal, 826, 196

\bibitem[{Pritchett \& Mozer(2009)}]{pritchett2009asymmetric}
Pritchett, P., \& Mozer, F. 2009, Journal of Geophysical Research. Space
  Physics, 114

\bibitem[{Roytershteyn {et~al.}(2015)Roytershteyn, Karimabadi, \&
  Roberts}]{roytershteyn2015generation}
Roytershteyn, V., Karimabadi, H., \& Roberts, A. 2015, Philosophical
  Transactions of the Royal Society A: Mathematical, Physical and Engineering
  Sciences, 373, 20140151

\bibitem[{Russell {et~al.}(1987)Russell, Riedler, Schwingenschuh, \&
  Yeroshenko}]{russell1987mirror}
Russell, C., Riedler, W., Schwingenschuh, K., \& Yeroshenko, Y. 1987,
  Geophysical research letters, 14, 644

\bibitem[{Shoji {et~al.}(2012)Shoji, Omura, \& Lee}]{shoji2012multidimensional}
Shoji, M., Omura, Y., \& Lee, L.-C. 2012, Journal of Geophysical Research:
  Space Physics, 117

\bibitem[{Soucek {et~al.}(2008)Soucek, Lucek, \&
  Dandouras}]{soucek2008properties}
Soucek, J., Lucek, E., \& Dandouras, I. 2008, Journal of Geophysical Research:
  Space Physics, 113

\bibitem[{Southwood \& Kivelson(1993)}]{southwood1993mirror}
Southwood, D.~J., \& Kivelson, M.~G. 1993, Journal of Geophysical Research:
  Space Physics, 98, 9181

\bibitem[{Stevens \& Kasper(2007)}]{stevens2007scale}
Stevens, M., \& Kasper, J. 2007, Journal of Geophysical Research: Space
  Physics, 112

\bibitem[{Sundberg {et~al.}(2015)Sundberg, Burgess, \&
  Haynes}]{sundberg2015properties}
Sundberg, T., Burgess, D., \& Haynes, C. 2015, Journal of Geophysical Research:
  Space Physics, 120, 2600

\bibitem[{Turner {et~al.}(1977)Turner, Burlaga, Ness, \&
  Lemaire}]{turner1977magnetic}
Turner, J., Burlaga, L., Ness, N., \& Lemaire, J. 1977, Journal of Geophysical
  Research, 82, 1921

\bibitem[{Valentini {et~al.}(2007)Valentini, Tr{\'a}vn{\'\i}{\v{c}}ek,
  Califano, Hellinger, \& Mangeney}]{valentini2007hybrid}
Valentini, F., Tr{\'a}vn{\'\i}{\v{c}}ek, P., Califano, F., Hellinger, P., \&
  Mangeney, A. 2007, Journal of Computational Physics, 225, 753

\bibitem[{Volwerk {et~al.}(2008)Volwerk, Zhang, Delva, V{\"o}r{\"o}s,
  Baumjohann, \& Glassmeier}]{volwerk2008mirror}
Volwerk, M., Zhang, T., Delva, M., {et~al.} 2008, Journal of Geophysical
  Research: Planets, 113

\bibitem[{Volwerk {et~al.}(2021)Volwerk, Mautner, Wedlund, Goetz, Plaschke,
  Karlsson, Schmid, Rojas-Castillo, Roberts, \& Varsani}]{angeo-39-239-2021}
Volwerk, M., Mautner, D., Wedlund, C.~S., {et~al.} 2021, Annales Geophysicae,
  39, 239, \dodoi{10.5194/angeo-39-239-2021}

\bibitem[{Winterhalter {et~al.}(2000)Winterhalter, Smith, Neugebauer,
  Goldstein, \& Tsurutani}]{winterhalter2000latitudinal}
Winterhalter, D., Smith, E.~J., Neugebauer, M., Goldstein, B.~E., \& Tsurutani,
  B.~T. 2000, Geophysical Research Letters, 27, 1615

\bibitem[{Yu {et~al.}(2021)Yu, Huang, Yuan, Jiang, Xiong, Xu, Wei, Zhang, \&
  Zhang}]{yu2021characteristics}
Yu, L., Huang, S., Yuan, Z., {et~al.} 2021, The Astrophysical Journal, 908, 56

\end{thebibliography}
\bibliographystyle{aasjournal}

\end{document}